# LOCAL SPECTRAL ATTENTION FOR FULL-BAND SPEECH ENHANCEMENT

Zhongshu Hou[1,2,3], Qinwen Hu[1,2,3], Kai Chen[1,2,3], Jing Lu[1,2,3]

[1] Key Laboratory of Modern Acoustics, Nanjing University, Nanjing 210093, China
[2] NJU-Horizon Intelligent Audio Lab, Horizon Robotics, Beijing 100094, China
[3] Nanjing Institute of Advanced Artificial Intelligence, Nanjing 210014, China

{zhongshu.hou, qinwen.hu}@smail.nju.edu.cn, {chenkai, lujing}@nju.edu.cn

## ABSTRACT

Attention mechanism has been widely utilized in speech enhancement (SE) because theoretically it can effectively model the inherent connection of signal both in time domain and spectrum domain. Usually the span of attention is limited in time domain while the attention in frequency domain spans the whole frequency range. In this paper, we notice that the attention over the whole frequency range hampers the inference for full-band SE and possibly leads to excessive residual noise. To alleviate this problem, we introduce local spectral attention (LSA) into full-band SE model by limiting the span of attention. The ablation test on the state-of-the-art (SOTA) full-band SE model reveals that the local frequency attention can effectively improve overall performance. The improved model achieves the best objective score on the full-band VoiceBank+DEMAND set.

## 1. INTRODUCTION

Speech enhancement (SE) is important in telecommunication systems, modern automatic speech recognition (ASR) and hearing aid devices [1,2], aiming at improving the perceptual quality and intelligibility as well as speech recognition performance of speech signals degraded by background disturbances. Data-driven deep neural network (DNN) based SE methods have recently shown significantly better performance over traditional rule-based signal processing approaches, especially under low signal-to-noise ratio (SNR) conditions with highly non-stationary noise [3].

Time-domain DNN based approaches directly enhance raw speech waveforms through the encoder-enhancer-decoder framework [4-6], whereas the majority of SE methods operate in the time-frequency (T-F) domain to estimate a mask between clean and distorted spectrum [7,8], or to directly predict the real and imaginary parts of the target clean complex spectrum from the noisy speech [9,10]. Different network structures have been used for temporal and spectral feature extraction in T-F models, including convolutional neural networks (CNN) [9-15], recurrent neural networks (RNN) [14-18], self-attention (SA) mechanisms [19-26].

Recently, SA-based models have shown their superiority in SE tasks, achieving state-of-the-art performances in both wideband (sampled at 16kHz) [25] (CMGAN) and full-band (sampled at 48kHz) [26] (MTFAA) scenarios. Temporal attention efficiently models the time dependencies of speech in a limited range, either by designating the number of attended frames for online processing or restricting the maximal length of the training sequence for offline processing, while the frequency-wise attention always acts on the whole frequency range, calculating attention weights globally. In this paper, we notice that the global spectral attention hinders the inference for full-band SE and probably leads to excessive residual noise. It may result from the feature discrepancy between different bands of speech, i.e., the structural harmonics in low bands and the almost randomly distributed components in high bands. Such weak similarity may produce large redundancy in attention mechanisms, for which SE models fail to focus on crucial characteristics, and thus result in suboptimal performance.

Local attention mechanism in natural language processing was introduced in [27], focusing on a subset of source words at a time to avoid expensive computation for effective machine translation. We find its potential to facilitate spectrum modeling with limited span at a certain frequency band. In this paper, we propose a local spectral attention (LSA) mechanism that only looks at adjacent bands at a certain frequency to alleviate the problem of full-band attention. We conduct experiments based on MTFAA model and dual-path attention-recurrent network (DPARN) [20] to demonstrate the advantage of the proposed LSA mechanism and achieve state-of-the-art (SOTA) results on the public full-band VoiceBank+DEMAND [28] dataset with the improved MTFAA.

## 2. METHOD DESCRIPTION

### 2.1. Spectral local attention

Let $\mathcal{X} \in \mathbb{R}^{T \times F \times 2}$ denote the input spectrogram to a T-F model with frequency-wise attention, where $T$ and $F$ denote the time and frequency dimensions, respectively. Then the key, query and score tensors passed to a conventional spectral attention module can be described as $\mathcal{Q}_F, \mathcal{K}_F, \mathcal{V}_F \in \mathbb{R}^{T \times F' \times C}$, where $F'$ is the frequency-wise dimension after down-sampling or up-sampling by CNNs and $C$ is the number of channels. Then the conventional attention output $\mathcal{A}_F \in \mathbb{R}^{T \times F' \times C}$ is formulated as

$$\mathbf{A}_{F,t} = \text{softmax}\left(\frac{\mathbf{Q}_{F,t}\mathbf{K}_{F,t}^{\mathrm{T}}}{\sqrt{F' \times C}}\right)\mathbf{V}_{F,t}, \quad (1)$$

$$\mathcal{A}_F = \text{concat}(\mathbf{A}_{F,1}, \mathbf{A}_{F,2}, \ldots, \mathbf{A}_{F,T}), \quad (2)$$

where $\mathbf{Q}_{F,t}, \mathbf{K}_{F,t}, \mathbf{V}_{F,t} \in \mathbb{R}^{F' \times C}$ represent key, query and score matrices at frame $t$, concat(∗) indicates concatenating the output attention matrix along the temporal dimension. The similarity matrix computed from the dot product of $\mathbf{Q}_{F,t}$ and $\mathbf{K}_{F,t}^{\mathrm{T}}$ forms the weights of the attention mechanism [19], which reflects the correlation among overall frequency bands. We consider that long-range spectrum correlation estimation may mislead the model to use insignificant information due to the weak similarity in feature distribution between low- and high- frequency bands, especially in full-band scenarios. For effective local spectrum dependency extraction, a local attention mask $\mathbf{M}_F \in \mathbb{R}^{F' \times F'}$ is applied to the similarity matrix, defined as

$$M_{F,i,j} = \begin{cases} 0, & |i - j| \leq N_l \\ -\infty, & |i - j| > N_l \end{cases}, \quad (3)$$

where $M_{F,i,j}$ denotes an entry of $\mathbf{M}_F$ at the $i^{th}$ row and the $j^{th}$ column, $N_l \leq F'$ is the local interaction dimension. Hence the proposed LSA output $\tilde{\mathcal{A}}_F$ can be calculated as

$$\tilde{\mathbf{A}}_{F,t} = \text{softmax}\left(\frac{\mathbf{Q}_{F,t}\mathbf{K}_{F,t}^{\mathrm{T}}}{\sqrt{F' \times C}} + \mathbf{M}_F\right)\mathbf{V}_{F,t}, \quad (4)$$

$$\tilde{\mathcal{A}}_F = \text{concat}(\tilde{\mathbf{A}}_{F,1}, \tilde{\mathbf{A}}_{F,2}, \ldots, \tilde{\mathbf{A}}_{F,T}). \quad (5)$$

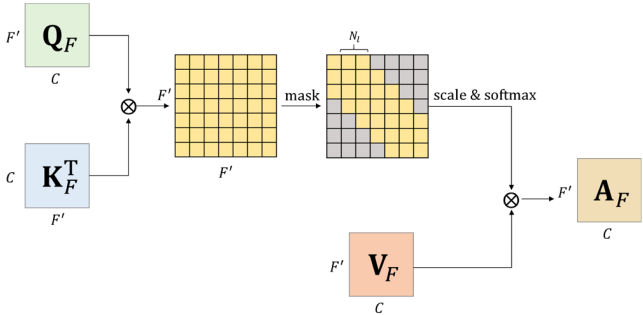

**Fig. 1.** The visualization of proposed LSA mechanism.

The visualization of the proposed LSA mechanism is shown in Fig. 1. In real network implementation, the $-\infty$ of $\mathbf{M}_F$ can be set as large negative value, such as -1e9. The source code of the proposed method is available at *https://github.com/ZhongshuHou/LSA*.

### 2.2. Model architecture

MTFAA network is the best full-band model in the 4th DNS challenge in ICASSP 2022 [26], which utilizes frequency-wise self-attention for spectrum modeling. It includes a phase encoder module to map complex spectral features to real components. Then the real spectral inputs are mapped according to ERB scale to reduce redundancy at high frequencies in band merging block. The band splitting is the inverse process of the band merging. The main-net consists of several similar processing units, where frequency down-sampling (FD) or up-sampling (FU) is implemented with 2-dimensional convolution. The TF-convolution module uses dilated convolution enlarging the receptive filed along the time domain and the axial self-attention (ASA) calculates attention matrices along the time and frequency axis for T-F feature modeling. In the mask estimating and applying procedure, deep-filter mask is applied to the magnitude spectrum in the first stage and the subsequent complex mask refines both the magnitude and phase spectrum. The detailed description on MTFAA can be found in [26]. In the original ASA block, attention operates along frequency in a global manner. We replace the frequency-wise global attention in ASA with the proposed LSA mechanism and carry out an ablation study on model performance in Sec. 3.

To further validate the efficacy of LSA, we also test its performance on another full-band SE model dubbed DPARN, which originates from DPCRN (rank 3rd in DNS-3) [17] but replaces the intra RNN module with multi-head self-attention for spectrum modeling.

### 2.3. Training targets and loss function

Power compress loss [29] functions are utilized to better process the information in low power T-F points:

$$\mathcal{L}_{RI}(\tilde{S}, S) = \left\|S_{real}^{\mathcal{C}} - \tilde{S}_{real}^{\mathcal{C}}\right\|_F^2 + \left\|S_{imag}^{\mathcal{C}} - \tilde{S}_{imag}^{\mathcal{C}}\right\|_F^2, \quad (6)$$

$$\mathcal{L}_{Mag}(\tilde{S}, S) = \left\||S|^\gamma - |\tilde{S}|^\gamma\right\|_F^2, \quad (7)$$

$$\mathcal{L}_{Ovrl}(\tilde{S}, S) = \mathcal{L}_{Mag}(\tilde{S}, S) + \mathcal{L}_{RI}(\tilde{S}, S), \quad (8)$$

$$S_{real}^{\mathcal{C}} = |S|^\gamma \cos\theta_S, \quad S_{imag}^{\mathcal{C}} = |S|^\gamma \sin\theta_S, \quad (9)$$

where $S$ and $\tilde{S}$ denote target clean and the enhanced spectrogram in the T-F domain, respectively, $\gamma$ refers to the compression parameter, superscript $\mathcal{C}$ denotes the power compressed pattern, $\theta_S$ denotes the phase angle of complex spectrogram, and $||\cdot||_F$ refers to the Frobenius norm of the matrix. Both the MTFAA network and DPARN are optimized with the overall loss function $\mathcal{L}_{Ovrl}(\tilde{S}, S)$.

## 3. EXPERIMENTS

### 3.1. Datasets

We demonstrate the effectiveness of our proposed method on the commonly used publicly available full-band Voice-Bank+DEMAND dataset [28]. The clean clips are selected from the VoiceBank corpus [30] with 28 speakers in the training set and 2 unseen speakers in the test set. The training set includes 11,572 utterances and the test set contains 872 utterances. In the training set, the clean utterances are mixed with background noise (8 types of real noise recordings from DEMAND database [31] and 2 artificial noise types) at SNRs of 0 dB, 5 dB, 10 dB and 15 dB. For the test set, the clean utterances are added with 5 unseen noise types from the DEMAND database at SNRs of 2.5 dB, 7.5dB, 12.5 dB and 17.5 dB. The selected noise types are mostly challenging, e.g., domestic noises (living room and kitchen), public space noises (restaurant, cafeteria and office) and transportation/street noises (car, bus, metro, busy traffic, subway station and public square). All utterances are sampled at 48 kHz in our experiments.

### 3.2. Parameter setup and training strategy

The window size and hop length of short time Fourier transformation (STFT) are 32ms and 8ms, respectively. The discrete Fourier transformation length is 1536 and the hanning widow is used for overlap-add waveform reconstruction. The local interaction dimension $N_l$ in MTFAA is set to [8, 4, 2, 2, 2, 4, 8, 16] at each ASA module, respectively, and the $N_l$ in DPARN is set to 64 at each multi-head self-attention block. Other parameter setups of MTFAA and DPARN can be found in [26] and [20], respectively.

The batch size in our training is 4 and the temporal length of input audio is 4s. Warmup strategy [19] is critical in training self-attention based model, where the learning rate $\alpha$ is updated with the rule: $\alpha = \frac{1}{\sqrt{C}} \times \min\left(\frac{1}{\sqrt{\varphi}}, \frac{\varphi}{\sqrt{\Psi^3}}\right)$, with $C = 64$, warmup steps $\Psi = 3000$ and $\varphi$ denoting the training step. We train the model by the warmup-based Adam optimizer with $\beta_1 = 0.9$, $\beta_2 = 0.98$, $\epsilon = 10^{-9}$. The compression parameter $\gamma$ is $\frac{1}{3}$.

### 3.3. Ablation study and evaluation metrics

To verify the efficacy of the proposed LSA over the conventional global attention mechanism, we carry out the ablation study on both the casual MTFAA and DPARN models, where causality is determined by the padded dilated convolution, masked attention and unidirectional RNN along time dimension. For extensive performance evaluation, we also compare the modified MTFAA model with previous full-band SOTA approaches on VoiceBank+DEMAND test set.

We choose a set of commonly used speech metrics to evaluate the comprehensive performance, i.e., perceptual evaluation of speech quality (PESQ) [32], short-time objective intelligibility (STOI) [33], scale-invariant signal distortion ratio (SiSDR) and composite mean opinion score (MOS) based metrics [34]: MOS prediction of the signal distortion (CSIG), MOS prediction of the intrusiveness of background noise (CBAK) and MOS prediction of the overall effect (COVL). Higher values indicate better performance for all metrics. Due to the lack of PESQ and composite MOS evaluation for full-band speech, we down-sample the enhanced results to 16 kHz (wideband) and measure the scores, while STOI and SiSDR are measured on full-band results.

### 3.4. Experimental results and analysis

#### 3.4.1. Ablation study

Ablation study results are presented in Table 1. The proposed LSA improves the enhancement performance of both the casual DPARN and MTFAA models in terms of all objective metrics. Further investigation of the enhanced signals reveals that the global attention in frequency domain is more likely to produce excessive residual noise especially in non-speech segments, while this problem can be effectively alleviated by the proposed LSA. Two typical examples are shown in Fig. 2, where the benefit of LSA can be clearly seen. A possible explanation is that the local attention may better exploit the speech pattern in spectrum and more effectively discriminate speech and noise components especially in low-SNR environments.

**Table 1.** Ablation study results

| Config. | | Wideband Metrics | | | | Full-band Metrics | |
|---|---|---|---|---|---|---|---|
| Model | LSA | PESQ | CSIG | CBAK | COVL | STOI(%) | SiSDR(dB) |
| MTFAA | X | 3.13 | 4.33 | 3.54 | 3.75 | 94.6 | 17.7 |
| | √ | **3.16** | **4.35** | **3.61** | **3.78** | **94.7** | **18.8** |
| DPARN | X | 2.92 | 4.26 | 3.61 | 3.65 | 94.2 | 18.3 |
| | √ | **2.96** | **4.29** | **3.63** | **3.68** | 94.2 | **18.7** |

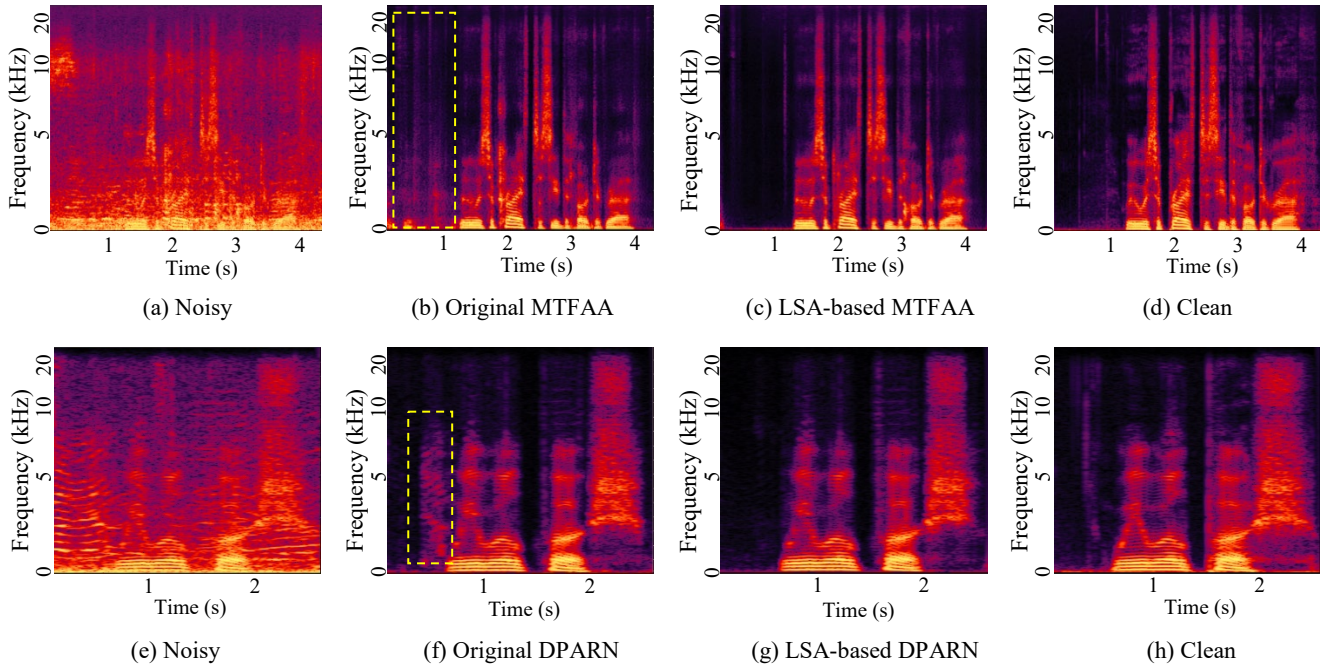

**Fig. 2.** Typical spectrograms. (a,e) Noisy speech, enhanced speech by (b,f) original non-LSA networks and (c,g) LSA-based networks, (d,h) Clean reference speech

**Table 2.** Objective metric results on VoiceBank+DEMAND test set. "-" indicates unreported scores of original papers

| Models | Year | Param.(M) | Wideband Metrics | | | | Full-band Metrics |
| --- | --- | --- | --- | --- | --- | --- | --- |
| | | | PESQ | CSIG | CBAK | COVL | STOI(%) |
| Noisy | - | - | 1.97 | 3.34 | 2.44 | 2.63 | 92.1 |
| S-DCCRN[35] | 2022 | 2.34 | 2.84 | 4.03 | 3.43 | 2.97 | 94.0 |
| FullSubNet+[36] | 2022 | 8.67 | 2.88 | 3.86 | 3.42 | 3.57 | 94.0 |
| GaGNet[37] | 2022 | 5.95 | 2.94 | 4.26 | 3.45 | 3.59 | - |
| DMF-Net[38] | 2022 | 7.84 | 2.97 | 4.26 | 3.52 | 3.62 | 94.4 |
| SF-Net[13] | 2022 | 6.98 | 3.02 | 4.36 | 3.54 | 3.67 | 94.5 |
| DeepFilterNet2[39] | 2022 | 2.31 | 3.08 | 4.30 | 3.40 | 3.70 | 94.3 |
| **MTFAA (Cau., LSA)** | 2022 | 1.50 | **3.16** | **4.35** | **3.61** | **3.78** | **94.7** |

*3.4.2. Comparison with full-band SOTA methods*

We compare the modified MTFAA model with previous full-band SOTA methods on VoiceBank+DEMAND dataset and the results are listed in Table 2. One can find that the modified causal MTFAA model outperforms previous SOTA methods on all evaluation metrics and achieves a 3.16 score in PESQ and an STOI value close to 95%. Furthermore, LSA mechanism results in fewer parameters and less computational complexity due to the local processing in frequency domain.

## 4. CONCLUSIONS

In this paper, we propose a local spectral attention mechanism for attention-based full-band speech enhancement models, which facilitates more effective extraction of spectral information. We merge the proposed LSA with full-band models, including DPARN and MTFAA, and validate its better performance in recovering speech with less residual noise. The MTFAA with LSA also achieves the best performance in VoiceBank+DEMAND test set. As a side benefit, the proposed method also leads to smaller models with less computation burden.

## 5. ACKNOWLEDGEMENTS

This work was supported by the National Natural Science Foundation of China (Grants No. 11874219 and No. 12274221).

# 6. REFERENCES

[1] F. Weninger and H. Erdogan, "Speech enhancement with lstm recurrent neural networks and its application to noise-robust asr," in *International conference on latent variable analysis and signal separation*. Springer, 2015, pp. 91–99.

[2] J. L. Desjardins and K. A. Doherty, "The effect of hearing aid noise reduction on listening effort in hearing-impaired adults," *Ear and Hearing*, vol. 35, pp. 600–610, 2014.

[3] DeLiang Wang and Jitong Chen, "Supervised speech separation based on deep learning: An overview," *IEEE/ACM Transactions on Audio, Speech, and Language Processing*, vol. 26, no. 10, pp. 1702–1726, 2018.

[4] Yi Luo and N. Mesgarani, "Conv-tasnet: Surpassing ideal time–frequency magnitude masking for speech separation," *IEEE/ACM transactions on audio, speech, and language processing*, vol. 27, no. 8, pp. 1256–1266, 2019.

[5] A. Defossez, G. Synnaeve, and Y. Adi, "Real Time Speech Enhancement in the Waveform Domain," in *Interspeech*, 2020, pp. 3291–3295.

[6] E. Kim and H. Seo, "Se-conformer: Time-domain speech enhancement using conformer.," in *Interspeech*, 2021, pp. 2736–2740.

[7] D. S. Williamson, Yuxuan Wang and DeLiang Wang, "Complex ratio masking for monaural speech separation," *IEEE/ACM transactions on audio, speech, and language processing*, vol. 24, no. 3, pp. 483–492, 2015.

[8] C. Hummersone, T. Stokes, and T. Brookes, "On the ideal ratio mask as the goal of computational auditory scene analysis," in *Blind source separation*, pp. 349–368. Springer, 2014

[9] Ke Tan and DeLiang Wang, "A convolutional recurrent neural network for real-time speech enhancement.," in *Interspeech*, 2018, vol. 2018, pp. 3229–3233.

[10] Andong Li, Wenzhe Liu, and Xiaoxue Luo, "A Simultaneous Denoising and Dereverberation Framework with Target Decoupling," in *Interspeech*, 2021, pp. 2801–2805.

[11] A. Pandey and DeLiang Wang, "Tcnn: Temporal convolutional neural network for real-time speech enhancement in the time domain," in *ICASSP*. IEEE, 2019, pp. 6875–6879.

[12] Yang Xian, Yang Sun, "Convolutional fusion network for monaural speech enhancement," *Neural Networks*, vol. 143, pp. 97–107, 2021.

[13] Guochen Yu, Andong Li, "Optimizing shoulder to shoulder: A coordinated sub-band fusion model for real-time full-band speech enhancement," *arXiv preprint arXiv:2203.16033*, 2022.

[14] Ke Tan and DeLiang Wang, "Complex Spectral Mapping with a Convolutional Recurrent Network for Monaural Speech Enhancement," in *ICASSP*. IEEE, 2019, pp. 6865–6869.

[15] Yanxin Hu, Yun Liu, "Dccrn: Deep complex convolution recurrent network for phase-aware speech enhancement," *arXiv preprint arXiv:2008.00264*, 2020.

[16] Zhong-Qiu Wang, S. Cornell, "Tf-gridnet: Making time-frequency domain models great again for monaural speaker separation," *CoRR*, vol. abs/2209.03952, 2022.

[17] Xiaohuai Le, Hongsheng Chen, "DPCRN: Dual-Path Convolution Recurrent Network for Single Channel Speech Enhancement," in *Interspeech*, 2021, pp. 2811–2815.

[18] Shengkui Zhao, Bin Ma, "Frcrn: Boosting feature representation using frequency recurrence for monaural speech enhancement," in *ICASSP*. IEEE, 2022, pp. 9281–9285.

[19] A. Nicolson and K. K. Paliwal, "Masked multi-head self-attention for causal speech enhancement," *Speech Communication*, vol. 125, pp. 80–96, 2020.

[20] Qinwen Hu, Zhongshu Hou, "A light-weight full-band speech enhancement model," *ArXiv*, vol. abs/2206.14524, 2022.

[21] C. Subakan, M. Ravanelli, "Attention is all you need in speech separation," in *ICASSP*. IEEE, 2021, pp. 21–25.

[22] A. Pandey and DeLiang Wang, "Dual-path self-attention rnn for real-time speech enhancement," *arXiv preprint arXiv:2010.12713*, 2020.

[23] A. Pandey, Buye Xu, "Tparn: Triple-path attentive recurrent network for time-domain multi-channel speech enhancement," in *ICASSP*. IEEE, 2022, pp. 6497–6501.

[24] Jngjing Chen, Qirong Mao, and Dong Liu, "Dual-path transformer network: Direct context-aware modeling for end-to-end monaural speech separation," *arXiv preprint arXiv:2007.13975*, 2020.

[25] S. Abdulatif, Ruizhe Cao, and Bin Yang, "Cmgan: Conformer-based metric-gan for monaural speech enhancement," *arXiv preprint arXiv:2209.11112*, 2022.

[26] Guochang Zhang, Libiao Yu, Chunliang Wang, and Jianqiang Wei, "Multi-scale temporal frequency convolutional network with axial attention for speech enhancement," in *ICASSP*. IEEE, 2022, pp. 9122–9126.

[27] T. Luong, H. Pham, and C. D. Manning, "Effective approaches to attention-based neural machine translation," *arXiv preprint arXiv:1508.04025*, 2015.

[28] C. V. Botinhao, Xin Wang, "Investigating rnn-based speech enhancement methods for noise-robust text-to-speech," in *SSW*, 2016.

[29] Andong Li, Chengshi Zheng, "On the importance of power compression and phase estimation in monaural speech dereverberation," *JASA Express Letters*, vol. 1, no. 1, pp. 014802, 2021.

[30] C. Veaux, J. Yamagishi, and S. King, "The voice bank corpus: Design, collection and data analysis of a large regional accent speech database," in *international conference oriental COCOSDA held jointly with 2013 conference on Asian spoken language research and evaluation*. IEEE, 2013, pp. 1–4.

[31] J. Thiemann, N. Ito, and E. Vincent, "The diverse environments multi-channel acoustic noise database (demand): A database of multichannel environmental noise recordings," *Journal of the Acoustical Society of America*, vol. 133, pp. 3591–3591, 2013.

[32] A. W. Rix, J. G. Beerends, and M. P. Hollier, "Perceptual evaluation of speech quality (pesq)-a new method for speech quality assessment of telephone networks and codecs," in *ICASSP*. IEEE, 2001, vol. 2, pp. 749–752.

[33] C. H. Taal, R. C. Hendriks, and R. Heusdens, "A short-time objective intelligibility measure for time-frequency weighted noisy speech," in *ICASSP*. IEEE, 2010, pp. 4214– 4217.

[34] Yi Hu and P. C. Loizou, "Evaluation of objective quality measures for speech enhancement," *IEEE Transactions on audio, speech, and language processing*, vol. 16, no. 1, pp. 229–238, 2007.

[35] Shubo Lv, Yihui Fu, "S-dccrn: Super wide band dccrn with learnable complex feature for speech enhancement," in *ICASSP*. IEEE, 2022, pp. 7767–7771.

[36] Jun Chen, Zilin Wang, "Fullsubnet+: Channel attention fullsubnet with complex spectrograms for speech enhancement," in *ICASSP*. IEEE, 2022, pp. 7857–7861.

[37] Andong Li, Chengshi Zheng, "Glance and gaze: A collaborative learning framework for single-channel speech enhancement," *Applied Acoustics*, vol. 187, pp. 108499, 2022.

[38] Guochen Yu, Yuansheng Guan, "Dmf-net: A decoupling-style multi-band fusion model for real-time full-band speech enhancement," *arXiv preprint arXiv:2203.00472*, 2022

[39] H. Schroter, T. Rosenkranz, "Deepfilternet2: Towards real-time speech enhancement on embedded devices for full-band audio," *arXiv preprint arXiv:2205.05474*, 2022.